\documentclass[english]{article}
\usepackage{geometry}
\usepackage[utf8]{inputenc}
\usepackage[T1]{fontenc}
\usepackage{babel}
\usepackage{amsmath}
\usepackage{graphicx}
\usepackage{fancyhdr}
\usepackage{url}
\usepackage{xcolor}
\usepackage{longtable}
\usepackage{makecell}
\usepackage{multirow}
\usepackage{gensymb}
\usepackage{cite}
\usepackage{tabularx}
\newcommand{\dataset}{OE62}

\usepackage{subcaption}
\begin{document}

\title{Atomic structures and orbital energies of 61,489 crystal-forming organic molecules}

\author{Annika Stuke\textsuperscript{1*}$^\dagger$, Christian Kunkel\textsuperscript{2}$^\dagger$, Dorothea Golze\textsuperscript{1}, Milica Todorovi\'{c}\textsuperscript{1},\\ Johannes T. Margraf\textsuperscript{2}, Karsten Reuter\textsuperscript{2}, Patrick Rinke\textsuperscript{1,2} \\ and Harald Oberhofer\textsuperscript{2}}

\maketitle
\thispagestyle{fancy}

1. Department of Applied Physics, Aalto University, P.O. Box 11100, Aalto FI-00076, Finland 2. Chair for Theoretical Chemistry and Catalysis Research Center, Technische Universit\"at M\"unchen, Lichtenbergstra\ss{}e 4, D-85747 Garching, Germany\\
 {*}corresponding author(s):
Annika Stuke (annika.stuke@aalto.fi), {$ ^\dagger$}both authors contributed equally to the present publication.

\begin{abstract}
Data science and machine learning in materials science require large datasets of technologically relevant molecules or materials. Currently, publicly available molecular datasets with realistic molecular geometries and spectral properties are rare. We here supply a diverse benchmark spectroscopy dataset of 61,489 molecules extracted from organic crystals in the Cambridge Structural Database (CSD), denoted \dataset. Molecular equilibrium geometries are reported at the Perdew-Burke-Ernzerhof (PBE) level of density functional theory (DFT) including van der Waals corrections for all 62k molecules. For these geometries, \dataset\ supplies total energies and orbital eigenvalues at the PBE and the PBE hybrid (PBE0) functional level of DFT for all 62k molecules in vacuum as well as at the PBE0 level for a subset of 30,876 molecules in (implicit) water. For 5,239 molecules in vacuum, the dataset provides quasiparticle energies computed with many-body perturbation theory in the $G_0W_0$ approximation with a PBE0 starting point (denoted GW5000 in analogy to the GW100 benchmark set (M. van Setten et al. J. Chem. Theory Comput. 12, 5076 (2016))). 
\end{abstract}

\section*{Background \& Summary}

Consistent and curated datasets have facilitated progress in the natural sciences. High-quality reference data sets were, for example, essential in the development of accurate computational methodology, in particular in quantum chemistry. With the rise of machine learning, datasets have increased in size and have transformed from reference status to a primary source of data for predictions \cite{BartokSciAdv2016, Schnet, Faber2017JCTC, SchuettNatComm2017,Tang2019JCP,stuke_chemical_2019,Gosh/etal:2019} and discovery \cite{Mansouri2018JACS,Meredig2014PhysRevB,Mayer2018ChemSci,Goldsmith2018AlChe,ShandizCompMatSci2016}.

In this article we present a new dataset for molecular spectroscopy applications. Spectroscopy is ubiquitous in science as one of the primary ways of determining a material's or molecule's properties. However, publicly available spectroscopic datasets for technologically relevant molecules are rare. Examples include a dataset of chemical shifts for structures taken from the CSD \cite{Paruzzo2018, Paruzzo_dataset}, the Harvard Clean Energy Project \cite{Hachmann2011} as well as the QM8 \cite{ramakrishnan_electronic_2015, ruddigkeit_2012} and QM9 \cite{ramakrishnan_quantum_2014} datasets. The QM8 database offers optical spectra computed with time-dependent density functional theory (TDDFT) for 22k organic molecules, while QM9, widely known as one of the standard benchmark sets for machine learning in chemistry, provides a variety of properties for 134k organic molecules computed with density functional theory (DFT) \cite{hedin_new_1965, kohn_nobel_1999}, including energy levels for the highest occupied and the lowest unoccupied molecular orbitals (HOMO and LUMO, respectively). Although QM8 and QM9 are of unprecedented size compared to previous, common benchmark sets in quantum chemistry of several hundred to thousands of molecules, they still contain only small molecules with restricted elemental diversity (H, C, N, O and F) and with simple bonding patterns \cite{stuke_chemical_2019}. They lack larger, more complex molecules with, e.g., extended heteroaromatic backbones and attached functional groups, as commonly targeted in organic synthesis\cite{Cabrele2016JOrgChem,Ponra2016RSCAdv} and applied in (opto-)electronic \cite{materials_odyssey_org_el,li2015organic,Ostroverkhova2016ChemRev, ostroverkhova2013handbook} or pharmaceutical research \cite{Ponra2016RSCAdv,silverman2014organic,TaylorOrgBiomolChem2016}.

\begin{figure}[h!]
      \centerline{\includegraphics[width=\textwidth]{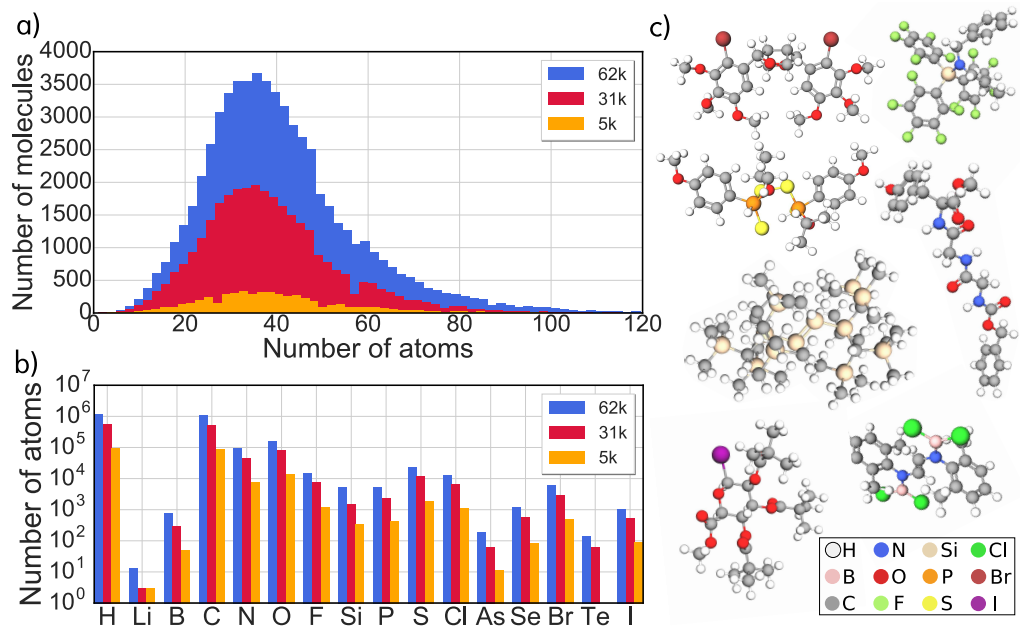}}
    \caption{Chemical space spanned by \dataset. (a) Molecular size distributions (including hydrogen atoms) for the \dataset\ dataset and its 31k and 5k subsets. (b) Distribution of the 16 different element types in the datasets. (c) Typical structures found in the 62k dataset, with chemical diversity arising from a rich combinatorial space of scaffold-functional group pairings: The dataset contains aliphatic molecules, as well as molecules with conjugated and complex aromatic backbones and diverse functional groups of technological relevance. The \texttt{refcode\_csd} identifiers of depicted molecules are (from left to right): \texttt{ZZTVO01}, \texttt{VOCMIK}, \texttt{FATVEC}, \texttt{WASVAN}, \texttt{BIDLUW}, \texttt{KETZAL}, \texttt{EHORAU}.} 
\label{fig:statistics}
\end{figure}

We have based the spectroscopic dataset presented in this article on a diverse collection of 64,725 organic crystals that were extracted from the Cambridge Structural Database (CSD) \cite{CSD} by Schober et al. \cite{Schober2016JPCL,SchoberDiss}. This \textit{64k dataset} of experimental crystal structures gathered from a variety of application areas was originally compiled to optimize the charge carrier mobility for applications in organic electronics. For our \dataset\ dataset, we  used 61,489 unique organic molecular structures, extracted from the respective organic crystals. All extracted geometries were then relaxed in the gas phase with density-functional theory (DFT). 

The molecules in \dataset\ cover a considerable part of chemical space, as illustrated in Figure~\ref{fig:statistics}. The dataset contains molecules with up to 174 (or 92 non-hydrogen) atoms and a diverse composition of 16 different elements. A large number of different scaffolds and functional groups are included, representing a multifaceted sample of the design space available in organic chemistry \cite{Schober2016JPCL, stuke_chemical_2019, MolLegoChemMater, KnowledgeDiscJMMO}.

\begin{figure}[h!]
      \centerline{\includegraphics[width=0.6\textwidth]{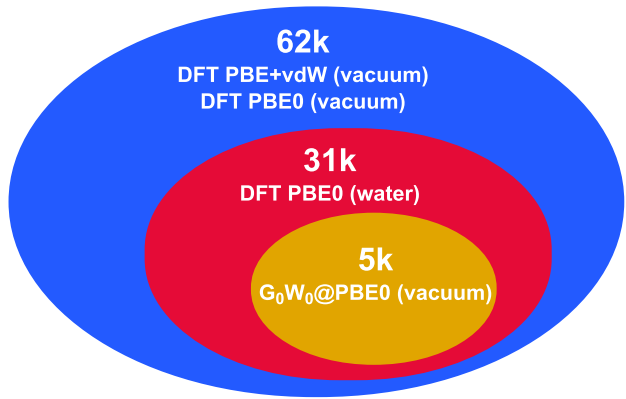}}
    \caption{\label{fig:sets} 
    Schematic overview of the three datasets and the applied computational methods. The 31k set includes all structures from the 5k set and the 62k all structures from the 31k and 5k sets.} 
    \label{fig:venn_diagram_sets}
\end{figure}

\begin{figure}[ht!]
\includegraphics[width=\textwidth]{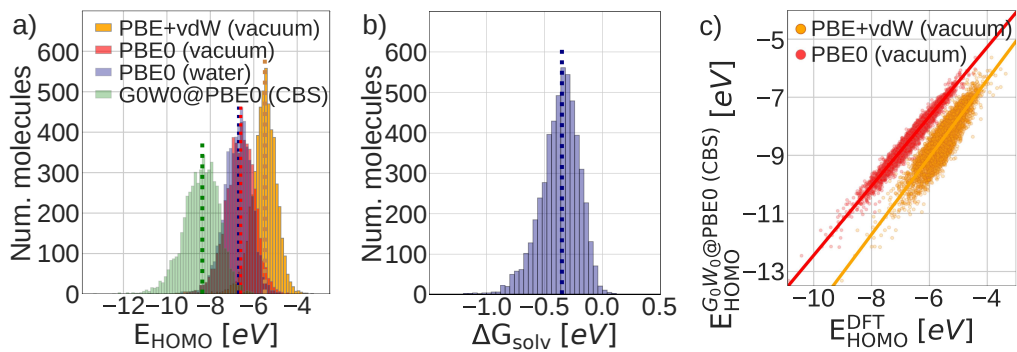}
    \caption{The GW5000 subset compared to the other (sub)sets in OE62. Panel (a) shows distributions of HOMO energies from $G_0W_0$@PBE0 (vacuum), PBE+vdW, PBE0 (vacuum) and PBE0 (water) computations. Panel (b) shows the distribution of solvation free energies $\rm{\Delta G_{solv}}=\rm{E_{tot}^{PBE0\;(water)}} -\rm{E_{tot}^{PBE0\;(vacuum)}}$. In a) and b), distribution medians are marked by dotted lines. Panel c) depicts a correlation plot for the approximately linear relationship between the $G_0W_0$@PBE0 CBS quasiparticle energies and the DFT HOMO energies (PBE and PBE0 in vacuum).}
\label{fig:differences_methods}
\end{figure}

To go into more detail, all molecules in \dataset\ are fully relaxed at the Perdew-Burker-Ernzerhof (PBE) \cite{PBE} level of DFT including Tkatchenko-Scheffler van der Waals (TS-vdW) corrections \cite{TS}. For these equilibrium structures, we then report molecular orbital energies at the PBE and PBE hybrid (PBE0) \cite{PBE0_2, PBE0_1} level, in the following referring to this part as 62k set. Partial charges and total energies for DFT-calculations are also included. In two subsets, randomly drawn to span more than half (31k) and more than 5000 (5k) of the molecular structures, we provide additional computational results: the influence of solvation -- in this case implicit water -- on the energy levels is addressed on the PBE0 level for a subset of 30,876 molecules. For the second subset of 5,239 molecules, we computed the quasi-particle energies with many-body perturbation theory in the $G_0W_0$ approximation \cite{hedin_new_1965,reining_the_2018,Golze2019} and extrapolated to the complete basis set (CBS) limit. Figure~\ref{fig:venn_diagram_sets} gives a schematic overview of the dataset nesting in \dataset\, while Table~\ref{tab:overview_subsets} lists computational settings and computed properties. Figure~\ref{fig:differences_methods}a) and b) illustrate the HOMO level and solvation free energy distributions of the 5k subset. 

We refer to the 5k subset of $G_0W_0$ quasiparticle energies as GW5000  in analogy to the GW100 benchmark set \cite{Setten2015}. GW100 was a landmark dataset of 100 atoms and molecules that for the first time demonstrated the high numerical accuracy of the computationally costly $G_0W_0$ approach. GW100 quickly became the standard reference for $GW$ code development and validation. The GW5000  subset in \dataset\ is of the same high numeric quality as GW100, but extends the set of reference molecules by a factor of 50. 
To illustrate the value of multi-level computational results we present a first, preliminary finding in Figure~\ref{fig:differences_methods}. Panel c) shows the correlation between the $G_0W_0$@PBE0  quasiparticle HOMO energies and the DFT HOMO eigenvalues for the GW5000  subset. The correlation is to first approximation linear with PBE0 having a lower variance than PBE. This linear relation (slope of 1.195 and intercept of -0.492 for PBE0) could now be used to predict $G_0W_0$ quasiparticle energies from the computationally cheaper PBE0 method without having to perform $G_0W_0$ calculations. Applying this linear correction to the PBE0 results yields quasiparticle energy predictions with a root mean square error (RMSE) of only 0.17 eV to the respective GW5000  values. 

Given the high-quality computational results from different levels of theory, the (subs)sets included in \dataset\ can be used to develop, train and evaluate machine learning algorithms, facilitating the search and discovery of diverse molecular structures with improved properties. In the following, we first describe the procedure used to compute molecular structures and properties, followed by a full description of the dataset format and content as well as by a validation of our DFT and $G_0W_0$ results. \dataset\ is freely available as a download from the Technical University of Munich. The input and output files of all calculations performed for \dataset\ can be downloaded from the Novel Materials Discovery (NOMAD) laboratory (\url{https://nomad-repository.eu}). 

\section*{Methods}

All crystal structures collected from the CSD for the \textit{64k dataset} are mono-molecular, i.e. they contain only a single type of molecule per unit-cell. A single molecular structure (conformer) from each crystal was extracted by a custom Python code \cite{Schober2016JPCL, SchoberDiss}. This \textit{64k dataset} of molecular structures provides the starting point for the dataset published here. A fraction of the crystals contained in the CSD have polymorphic forms or were added multiple times, coming e.g. from different experimental sources. Although they occur in different crystalline entries in the 64k dataset, the same molecular structure could enter our molecular database multiple times. First, the SMILES identifiers were computed for the \textit{64k dataset} \cite{SchoberDiss, Schober2016JPCL} from a combination of Open Babel \cite{Openbabel} (\url{www.openbabel.org}) and RDKit (\url{www.rdkit.org}) \cite{rdkit}.
We subsequently excluded all extracted molecules whose non-isomeric, canonical SMILES identifier occurred multiple times, keeping only one case each. Further, molecules with an odd number of electrons were removed. After these filtering steps 61,539 molecules remained. 

We relaxed the geometries of all molecules at the PBE$+$vdW level of theory, as implemented in the FHI-aims all-electron code \cite{BLUM20092175,ren_resolution_2012,AIMS2}. We chose the PBE+vdW functional for three reasons: 1) It is an all-purpose functional with a favorable accuracy/computational cost ratio that is implemented in all the major electronic structure codes. 2) We would like to stay consistent with  previous work \cite{Ropo_2016_amino_acids,stuke_chemical_2019}, in which PBE+vdW was also used for molecular structures optimization of large molecular data sets. 3) While there might be more accuracte semi-local functionals than PBE \cite{Mardirossian/Head-Gordon:2017}, the addition of vdW corrections makes PBE+vdW appropriate for organic compounds. For organic crystals, for which highly accurate, low-temperature experimental geometries are available, PBE+vdW yields excellent agreement with typical root-mean-squared deviations of only 0.005 - 0.01 {\AA} per atom \cite{marom_structure_2011,reilly_understanding_2013, hoja_first_2018}. 

Given that slightly differing bond assignments in the newly obtained low-energy geometries might change some of the molecular identifiers, we generated new InChI \cite{Heller2015} (’IUPAC International Identifier’) and canonical SMILES identifiers using Open Babel (Version 2.4.1 2016), and report these in our dataset. We then checked these representations for duplicates and concurrently removed them. In addition, 6 molecules were removed for which geometry optimization or single point calculations had failed. In total, 61,489 unique molecules remained, which form the basis of the \dataset\ set. 

From the \dataset\ set we generated two subsets: For the 31k subset we randomly picked  30,876 molecules. The same was done for the 5k set by randomly picking 5,239 molecules from the 31k subset with the additional constraint that the largest molecule should not exceed 100 atoms. The size and element distributions of all three sets are show in Figure~\ref{fig:statistics}. 

In the following we explain the data and additional subsets we created and provide the computational settings. All settings are also listed in Table~\ref{Table:methods}. 

\newcolumntype{L}[1]{>{\raggedright\arraybackslash}p{#1}}

\begin{center}
\begin{table}[h!]
\begingroup
\footnotesize
\renewcommand\arraystretch{4}
\begin{tabular}{|m{0.04\textwidth}|m{0.34\textwidth}|m{0.39\textwidth}|L{2cm}|}
\hline
\hline
\textbf{Set}  & \textbf{Method}  &  \textbf{Computed properties} & \textbf{Access to data records on NOMAD}\\
\hline
 62k
 & \makecell[l]{\textbf{DFT PBE $+$ vdW (vacuum)} \\ \textit{Tier2 basis set, tight settings}}
 & \makecell[l]{\textbullet~relaxed geometry\\ \textbullet~occupied \& unoccupied MO energies \\ \textbullet~total energy \\ \textbullet~Hirshfeld charges}  
 & \cite{nomad_pbe_vdw_part1, nomad_pbe_vdw_part2, nomad_pbe_vdw_part3, nomad_pbe_vdw_part4, nomad_pbe_vdw_part5, nomad_pbe_vdw_part6, nomad_pbe_vdw_part7}
\\
\hline
62k 
& \makecell[l]{\textbf{DFT PBE0 (vacuum)}\\ \textit{Tier2 basis set, tight settings}}
& \makecell[l]{\textbullet~geometry fixed at the PBE$+$vdW level \\ \textbullet~occupied \& unoccupied MO energies \\ \textbullet~total energy \\ \textbullet~Hirshfeld charges} 
& \cite{nomad_pbe0_vacuum}
\\
\hline
31k 
& \makecell[l]{\textbf{DFT PBE0 (water)} \\ \textit{Tier2 basis set, tight settings,}  \\ \textit{MPE implicit solvation}}
& \makecell[l]{\textbullet~geometry fixed at the PBE$+$vdW level \\ \textbullet~occupied \& unoccupied MO energies \\ \textbullet~total energy \\ \textbullet~Hirshfeld charges} 
& \cite{nomad_pbe0_water}
\\
\hline
5k
& \makecell[l]{\textbf{DFT PBE0 (vacuum)} \\ \textit{def2-TZVP \& def2-QZVP basis sets } \\ \textit{(see text), tight settings}}
& \makecell[l]{\textbullet~geometry fixed at the PBE$+$vdW level \\ \textbullet~occupied \& unoccupied MO energies \\ \textbullet~total energy} 
& \cite{nomad_tzvp}
\\
\hline 
5k
& \makecell[l]{\textbf{$\boldsymbol{G_0W_0}$@PBE0 (vacuum)} \\ \textit{def2-TZVP \& def2-QZVP basis sets } \\ \textit{(see text), tight settings}}
& \makecell[l]{\textbullet~geomety fixed at the PBE$+$vdW level \\ \textbullet~occupied \& unoccupied MO energies \\ \textbullet~CBS energies of occupied \& unoccu-\\ \hspace{0.15cm} pied MOs}  
& \cite{nomad_qzvp}
\\
\hline
\hline
\end{tabular}
\endgroup
\captionof{table}{\label{Table:methods}Overview of the data (sub)sets in \dataset: Applied computational method, resulting molecular properties and DOI-based references to the input and output files of corresponding calculations deposited in the NOMAD repository.}
\label{tab:overview_subsets}
\end{table}
\end{center}

\subsection*{62k set: DFT PBE $+$ vdW (vacuum)}
We pre-relaxed all molecular geometries at the PBE level of theory. For structure relaxation, we used the trust radius enhanced variant of the Broyden-Fletcher-Goldfarb-Shanno (BFGS) algorithm as implemented in FHI-aims with a maximum atomic residual force criterion of $f_{\rm{max}}$ < 0.01\:eV\,\r{A}$^{-1}$. The  electronic  wave  functions were expanded in  a Tier1 basis set at light integration settings \cite{BLUM20092175}. Since our database only contains closed-shell molecules, we performed spin-restricted DFT calculations. Dispersive forces were included in the geometry relaxations using the Tkatchenko-Scheffler (TS) \cite{TS} method, while relativistic effects were treated on the level of the atomic zero-order regular approximation (atomic ZORA)~\cite{BLUM20092175}. The DFT self-consistency cycle was treated as converged when changes of total energy, sum of eigenvalues and charge density were found below $10^{-6}$ eV, $10^{-3}$ eV and $10^{-5}$ e\,\r{A}$^{-3}$, respectively. Starting from these pre-relaxed structures, we obtained the final geometries by performing a new relaxation with Tier2 basis sets, tight integration settings and a convergence criterion of $f_{\rm{max}}$ < 0.001\:eV\,\r{A}$^{-1}$. The eigenvalues of the molecular states are then stored in our dataset alongside the molecular geometries. We refer to this part of the dataset as \textit{PBE$+$vdW (vacuum)}. 

\subsection*{62k set: DFT PBE0 (vacuum)}
Using the relaxed geometries obtained at the PBE$+$vdW (vacuum) level of theory, we further carried out single point calculations for all structures using the PBE0 hybrid functional. Computational settings as described before were used, employing again the Tier2 basis set with a tight integration grid. Note, that tabulated total energies obtained at this level also include the vdW contribution computed through the TS method, while "vdW" was dropped from the name to emphasize the single point character of these computations. We correspondingly refer to this set as \textit{PBE0 (vacuum)}.

\subsection*{31k subset: DFT PBE0 (water)}
To study the influence of solvation---here by water---on the PBE0 results, we performed calculations using the Multipole Expansion (MPE) implicit solvation method as implemented in FHI-aims \cite{MPE} for the 31k susbset. The MPE method facilitates an efficient treatment of the solvation effects on a solute, by using a continuum model of the solvent around it. In detail, the solute molecule is placed within a cavity with the dielectric permitivity of vacuum. The position of the cavity surface is determined by an iso-value $\rho_{\rm{iso}}$ of the solute's electronic density. Outside of this cavity the dielectric constant of water $\varepsilon _b$ = 78.36 was applied \cite{MPE}. 
The density isovalue $\rho_{iso}$ as well as the $\alpha$ and $\beta$ parameters for non-electrostatic contributions to the solvation free energy were taken from the published SPANC parameter-set \cite{MPE}. 

In the MPE method, the solvation cavity is discretized using a large number of points homogeneously distributed at the density iso-surface. Sampling of these points was achieved using an inexpensive pseudo-dynamical optimisation, allowing up to 1000 optimisation steps and removing the worst 0.1 \% of walkers at each neighbor-list update step\cite{MPE}, to account for the more complex molecules included in the 62k dataset. To obtain highly converged eigenvalues, we increased the reaction field- and polarization potential expansion orders $l_{\rm{max,R}}$ and $l_{\rm{max,O}}$ to 14 and 8, respectively, and the degree of overdetermination $d_{\rm{od}}$ to 16, keeping all other parameters at their default values \cite{MPE}. Note that the molecular geometries were not further relaxed in the presence of the water solvent. We kept the structures fixed at the PBE$+$vdW level. Tabulated total energies again include the vdW contribution obtained by the TS method. The resulting data is referred to as \textit{PBE0 (water)}. 

\subsection*{5k subset: $\boldsymbol{G_0W_0}$@PBE0 (vacuum)}
For the 5k subset, the relaxed PBE$+$vdW structures in vacuum were used as input for the $G_0W_0$~\cite{hedin_new_1965,Golze2019, Aryasetiawan1998} calculations, using the FHI-aims $G_0W_0$ implementation based on the analytic continuation \cite{ren_resolution_2012}. The PBE0 hybrid functional was used for the underlying DFT calculation ($G_0W_0$@PBE0) in combination with the atomic ZORA approximation.

In these $G_0W_0$ and PBE0 calculations, we employed the def2 triple-zeta valence plus polarization (def2-TZVP) and the def2 quadruple-zeta valence plus polarization (def2-QZVP) basis sets~\cite{weigend_balanced_2005}. The def2-TZVP and def2-QZVP basis sets are contracted Gaussian orbitals, treated numerically to be compliant with the numeric atom-centered orbital (NAO) technology in FHI-aims~\cite{BLUM20092175}. They are fully all-electron for all elements and do not contain effective core potentials. The def2 basis sets are available from the EMSL database \cite{feller_the_1996, schuchardt_basis_2007}, except for iodine (see Supplementary Information). Note that a basis set of def2-TZVP quality is not available for I and all def2-TZVP calculations for iodine-containing molecules were correspondingly performed with def2-QZVP for I and with def2-TZVP for all other elements.

Since $G_0W_0$ calculations converge slowly with respect to basis set size \cite{Golze2019}, we extrapolated the quasiparticle energies to the complete basis set (CBS) limit. Following the procedure for the GW100 benchmark set~\cite{Setten2015}, the extrapolated values are calculated from the def2-TZVP and def2-QZVP results by a linear regression against the inverse of the total number of basis functions (see Technical Validation).

The $G_0W_0$ self-energy elements were calculated for a set of imaginary frequencies $\{i\omega\}$ and then analytically continued to the real frequency axis using a Pad{\'e} approximant \cite{vidberg_solving_1977} with 16 parameters. The numerical integration along the imaginary frequency axis $\{i\omega'\}$ was performed using a modified Gauss-Legendre grid \cite{ren_resolution_2012} with 200 grid points. The same grid was employed for the set of frequencies $\{i\omega\}$, for which the self-energy is computed. The analytic continuation in combination with the Pad{\'e} model yields accurate results for valence states~\cite{Setten2015}, but is not reliable for core and semi-core states \cite{Golze2018}. Therefore, we included only occupied states with quasiparticle energies larger than -30~eV in the data set, see also Technical Validation for more details.

\subsection*{Code Availability}
All electronic structure data contained in this work was generated with the FHI-aims code \cite{BLUM20092175,ren_resolution_2012,AIMS2}. The code is available for a license fee from  \url{https://aimsclub.fhi-berlin.mpg.de/aims_obtaining_simple.php}. Parsing of outputs and data collection were performed with custom-made Python scripts, which will be available upon request. Finally, the published archive contains a tutorial detailing how to access the dataset.

\section*{Data Records}
The curated data for all 61,489 molecules is publicly available from two sources:

\begin{enumerate}
\item The dataset and related files can be freely downloaded from the media repository of the Technical University of Munich (mediaTUM) under \url{https://doi.org/10.14459/2019mp1507656} \cite{mediatum}. The dataset is provided as JSON output data of Pandas \cite{pandas} DataFrames. Within Python, these dataframes allow structured access to data in a tabular format, where each molecule is stored in a row of the dataframe, while the data is organized in columns. The content of the dataframe is summarized and explained in Table~\ref{Table:columns}. We also provide a tutorial file, which explains loading, filtering and data extraction from dataframes within Python. On mediaTUM, the dataset is distributed under a Creative Commons licence (
\url{https://creativecommons.org/licenses/by-sa/4.0/}).

\item The input and output files of all performed calculations can be downloaded from NOMAD. Due to the size of \dataset\, we provide an individual DOI for each applied computational method \cite{nomad_pbe_vdw_part1, nomad_pbe_vdw_part2, nomad_pbe_vdw_part3, nomad_pbe_vdw_part4, nomad_pbe_vdw_part5, nomad_pbe_vdw_part6, nomad_pbe_vdw_part7, nomad_pbe0_vacuum, nomad_pbe0_water, nomad_tzvp, nomad_qzvp}.
\end{enumerate}

\subsection*{Dataframe format}
We provide three dataframes: \texttt{df$\_$62k}, \texttt{df$\_$31k} and \texttt{df$\_$5k}. For each molecule in these dataframes, we provide three identifiers (\texttt{refcode$\_$csd}, \texttt{canonical$\_$smiles} and \texttt{inchi} in columns 1 to 3). In column 5, atomic coordinates of PBE $+$ vdW (vacuum) relaxed structures are stored as a string in a standard XYZ format (\texttt{xyz$\_$pbe$\_$relaxed}): The structure information contains a header line specifying the number of atoms $n_a$, an empty comment line and $n_a$ lines containing element type and relaxed atomic coordinates, one atom per line. The structure of all three dataframes is summarized in Table~\ref{Table:columns}.

The following list provides a brief overview over the three dataframes:
\begin{itemize}

\item Dataframe \texttt{df$\_$5k} includes 5,239 structures with results for all molecular properties in columns \textbf{5} to \textbf{29}. 

\item Dataframe \texttt{df$\_$31k} accommodates 30,876 structures, including all structures from \texttt{df$\_$5k}. $G_0W_0$@PBE0 results are only available for molecules from its 5k subset, while respective columns are left blank for the remaining molecules in \texttt{df$\_$31k}. 

\item Dataframe \texttt{df$\_$62k} contains all 61,489 structures, including all structures from \texttt{df$\_$31k} and \texttt{df$\_$5k}. PBE0 (water) results are only available for molecules from its 31k subset, while respective columns are left blank for the remaining molecules in \texttt{df$\_$62k}. The same applies for $G_0W_0$@PBE0 results for the structures from the 5k subset. The dataframe is ordered, such that the molecules included in the 5k subset are included first, while the remaining molecules of 31k and 62k subsets follow subsequently. This data structure facilitates the filtering of the dataframe by single lines of code, as shown in the tutorial.
\end{itemize} 

In addition, a spreadsheet file is provided in the distributed archive which contains the total energies of all atomic species of the dataset. They are computed for the respective levels of theory using similar computational settings, so that atomization energies for all molecules can be computed from the available molecular total energies. 

Finally, future updated versions of the dataset on mediaTUM will be distributed through the versioned DOI given above. In such cases, updated descriptions will be provided in the distributed archive alongside the dataset.

\newpage
\begin{center}
\begingroup
\scriptsize
\renewcommand\arraystretch{1.6}
\begin{longtable}{|p{0.3cm}|m{3.7cm}|p{0.45cm}|>{\raggedright\arraybackslash}p{2.7cm}|m{1.51cm}|m{3.8cm}|}
\hline
\hline
 \textbf{No.}  & \textbf{Column name} & \textbf{Unit} & \textbf{Method} & \textbf{Dataframes} & \textbf{Description}\\
\hline
\textbf{1} & \texttt{refcode\_csd} & --- & --- & 62k, 31k, 5k & CSD reference code, unique identifier for the crystal from which the molecule was extracted\\
\hline
\textbf{2} & \texttt{canonical\_smiles} & --- & Open Babel & 62k, 31k, 5k & \multirow{2}*{\makecell[l]{Molecular string representations \\ derived from DFT PBE$+$vdW \\relaxed geometries.}} \\
\textbf{3} & \texttt{inchi} & --- & Open Babel & 62k, 31k, 5k & \\
\hline
\textbf{4} & \texttt{number\_of\_atoms} & --- & --- & 62k, 31k, 5k & Number of atoms in the molecule\\
\hline
\textbf{5} & \texttt{xyz\_pbe\_relaxed} & \AA & PBE$+$vdW (vacuum) & 62k, 31k, 5k & String in XYZ-file format of DFT PBE$+$vdW relaxed geometry. Line 1 contains the number of atoms. Line 2 is empty. The remaining lines contain atomic type and coordinate (x,y,z).\\
\hline
\textbf{6} & \texttt{energies\_occ\_pbe} & $eV$ & PBE$+$vdW (vacuum) & 62k, 31k, 5k & \multirow{7}*{\makecell[l]{List of eigenvalues of occupied \\molecular Kohn-Sham orbitals. \\Given in ascending order, the \\last value is the HOMO energy.}}  \\
\textbf{7} & \texttt{energies\_occ\_pbe0\_vac\_tier2} & $eV$ & PBE0 (vacuum) & 62k, 31k, 5k & \\
\textbf{8} & \texttt{energies\_occ\_pbe0\_water} & $eV$& PBE0 (water) & 31k, 5k & \\
\textbf{9} & \texttt{energies\_occ\_pbe0\_vac\_tzvp} & $eV$ & PBE0 (vacuum) & 5k & \\
\textbf{10} & \texttt{energies\_occ\_pbe0\_vac\_qzvp} & $eV$ & PBE0 (vacuum) & 5k & \\
\textbf{11} & \texttt{energies\_occ\_gw\_tzvp} & $eV$ & $G_0W_0$@PBE0 (vacuum) & 5k & \\
\textbf{12} & \texttt{energies\_occ\_gw\_qzvp} & $eV$ & $G_0W_0$@PBE0 (vacuum) & 5k & \\
\hline
\textbf{13} & \texttt{cbs\_occ\_gw} & $eV$ & $G_0W_0$@PBE0 (vacuum) & 5k & List of CBS energies of occupied states computed from $G_0W_0$@PBE0 TZVP and QZVP energies from \textbf{10} and \textbf{11}. Same order as lists described above. \\
\hline
\textbf{14} & \texttt{energies\_unocc\_pbe} & $eV$ & PBE$+$vdW (vacuum)  & 62k, 31k, 5k & \multirow{9}*{\makecell[l]{List of eigenvalues of virtual \\(unoccupied) molecular Kohn-\\Sham orbitals. Given in ascen-\\ding order, the first value is the \\LUMO energy. Only virtual \\states below the vacuum level \\(i.e. with negative eigenvalue) \\are listed. If the LUMO energy \\is positive, only the LUMO \\energy is listed. If \textbf{20} has more \\negative eigenvalues than \textbf{19},\\ we also include positive eigen-\\values in \textbf{19} so that both lists\\ in \textbf{19} and \textbf{20} have equal length.}}\\
\textbf{15} & \texttt{energies\_unocc\_pbe0\_vac\_tier2} & $eV$ & PBE0 (vacuum) & 62k, 31k, 5k & \\
\textbf{16} & \texttt{energies\_unocc\_pbe0\_water} & $eV$ & PBE0 (water) & 31k, 5k & \\
\textbf{17} & \texttt{energies\_unocc\_pbe0\_vac\_tzvp} & $eV$ & PBE0 (vacuum)& 5k & \\
\textbf{18} & \texttt{energies\_unocc\_pbe0\_vac\_qzvp} & $eV$ & PBE0 (vacuum)& 5k & \\
\textbf{19} & \texttt{energies\_unocc\_gw\_tzvp} & $eV$ & $G_0W_0$@PBE0 (vacuum) & 5k & \\
\textbf{20} & \texttt{energies\_unocc\_gw\_qzvp} & $eV$ & $G_0W_0$@PBE0 (vacuum) & 5k & \\
 &  &  &  & & \\
\hline
\textbf{21} & \texttt{cbs\_unocc\_gw} & $eV$ & $G_0W_0$@PBE0 (vacuum) & 5k & List of CBS energies of unoccupied states computed from $G_0W_0$@PBE0 TZVP and QZVP energies from \textbf{19} and \textbf{20}. Same order as lists described above.  \\
\hline
\textbf{22} & \texttt{total\_energy\_pbe} & $eV$ & PBE$+$vdW (vacuum) & 62k, 31k, 5k & \multirow{5}*{\makecell[l]{Total energy of the DFT calcu-\\lations. Note, for consistency\\
with \textbf{22}, \textbf{23} 
and \textbf{24} also include\\ the vdW contribution to the\\ total energy. \\
\textbf{25} and \textbf{26} do not include it.}}\\
\textbf{23} & \texttt{total\_energy\_pbe0\_vac\_tier2} & $eV$ & PBE0 (vacuum)& 62k, 31k, 5k & \\
\textbf{24} & \texttt{total\_energy\_pbe0\_water} & $eV$ & PBE0 (water) & 31k, 5k & \\
\textbf{25} & \texttt{total\_energy\_pbe0\_vac\_tzvp} & $eV$ & PBE0 (vacuum) & 5k & \\
\textbf{26} & \texttt{total\_energy\_pbe0\_vac\_qzvp} & $eV$ & PBE0 (vacuum) & 5k & \\
\hline
\textbf{27} & \texttt{hirshfeld\_pbe} & $q_e$ & PBE$+$vdW (vacuum) & 62k, 31k, 5k & \multirow{3}*{\makecell[l]{List of Hirshfeld partial charges \\on atoms. Same order as atoms \\in \texttt{xyz\_pbe\_relaxed}.}}\\
\textbf{28} & \texttt{hirshfeld\_pbe0\_vac\_tier2} &  $q_e$  & PBE0 (vacuum) & 62k, 31k, 5k & \\
\textbf{29} & \texttt{hirshfeld\_pbe0\_water} & $q_e$ & PBE0 (water)& 31k, 5k & \\
\hline
\hline
\caption{\label{Table:columns}Dataframe structure of all three dataframes \texttt{df$\_$62k}, \texttt{df$\_$31k} and \texttt{df$\_$5k}. Columns 1 to 3 contain molecular identifiers. Columns 5 to 29 contain molecular properties computed at respective level of theory. All mentioned energies are given in eV. }
\label{tab:field_description}

\end{longtable}
\endgroup
\end{center}

\section*{Technical Validation}

\subsection*{Validation of relaxed geometries}

\begin{figure}[h!]
    \centerline{\includegraphics[width=\textwidth]{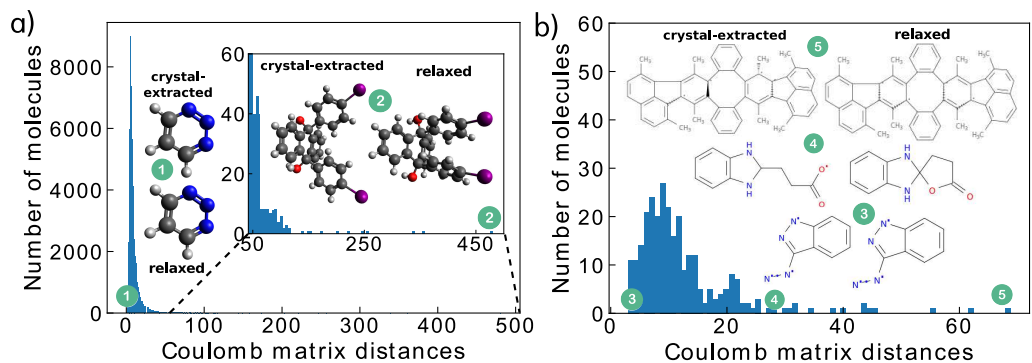}}
    \caption{Coulomb matrix distances between initial crystal geometries and PBE$+$vdW relaxed geometries. Panel a) shows the distribution of Coulomb matrix distances for all 62k molecules and panel b) shows the distribution of Coulomb matrix distances for the 284 cases that did not pass the consistency check. Two example molecules are shown in a) for short and large distances between Coulomb matrices (the \texttt{refcode\_csd} identifiers are \texttt{CILWUP} (1) and \texttt{ODAHUW} (2)). In b), 2D structures of three example molecules that failed the consistency check are shown (\texttt{DAZIND} (3), \texttt{YOMDUA} (4) and \texttt{FODBAC} (5)).}
\label{fig:cm_distances}
\end{figure}

To quantify the degree to which relaxation in vacuum changes the geometry of the structures compared to their crystalline form, we computed the distance between the two Coulomb matrices \cite{rupp20012_qm7,DScribe} of the original crystal geometry and the PBE$+$vdW relaxed geometry for each of the 62k molecules. The distribution of these Coulomb matrix distances is shown in Figure \ref{fig:cm_distances}a). Small distances signify small changes and large distances signify large differences. Most molecules exhibit only little changes in geometry during relaxation, where bond lengths are shifted by a small amount, as illustrated for the example of molecule 1. In some rare cases we find significant shifts in geometry caused by the environmental change from intermolecular interactions in the crystal to intramolecular interactions in vacuum, as shown for molecule 2. The crystal-extracted structure is shaped according to intermolecular van der Waals interactions that were present in the crystal. After relaxation, the intramolecular interactions cause a contraction of the molecular structure. 

To validate that the chemical integrity of the majority of the 62k molecules is preserved during the PBE$+$vdW relaxation, we perform a consistency check similarly to Ref.~\cite{ramakrishnan_quantum_2014}. We generated InChI strings from the relaxed PBE$+$vdW geometries and compare them to those obtained from the initial crystal-extracted cartesian coordinates. For 284 pairs, the two InChI strings did not match. Such mismatches can, for example, be caused by specifics in the implementation, in which Openbabel assigns different InChI strings to molecules with the same topology, possibly caused by changes in bond lengths, bond angles or dihedral angles. Examples are shown in Figure~\ref{fig:cm_distances}b) with molecule 3 exhibiting a small Coulomb matrix distance or molecule 5, which exhibits a large Coulomb matrix distance due to stronger relaxation. Here, stereoassignments change in the molecular structure, causing the different InChI-identifiers. Conversely, the mismatch can be also caused by changes in molecular topology during relaxation. This is the case for molecule 4, for which an intramolecular ring-closure takes place. Compared to 3,054 such inconsistencies found during the collection of the 134k molecules for the QM9 database \cite{ramakrishnan_quantum_2014}, the number of 284 found here is considerably small. The reason is most likely that our molecular starting geometries were derived from experimentally observed, well-resolved solid-form conformers.

\subsection*{\label{sec:extrapolation}Validation of DFT atomization and orbital energies}

\begin{center}
\begin{figure}[ht!]
\begin{center}
     \includegraphics[width=0.95\textwidth]{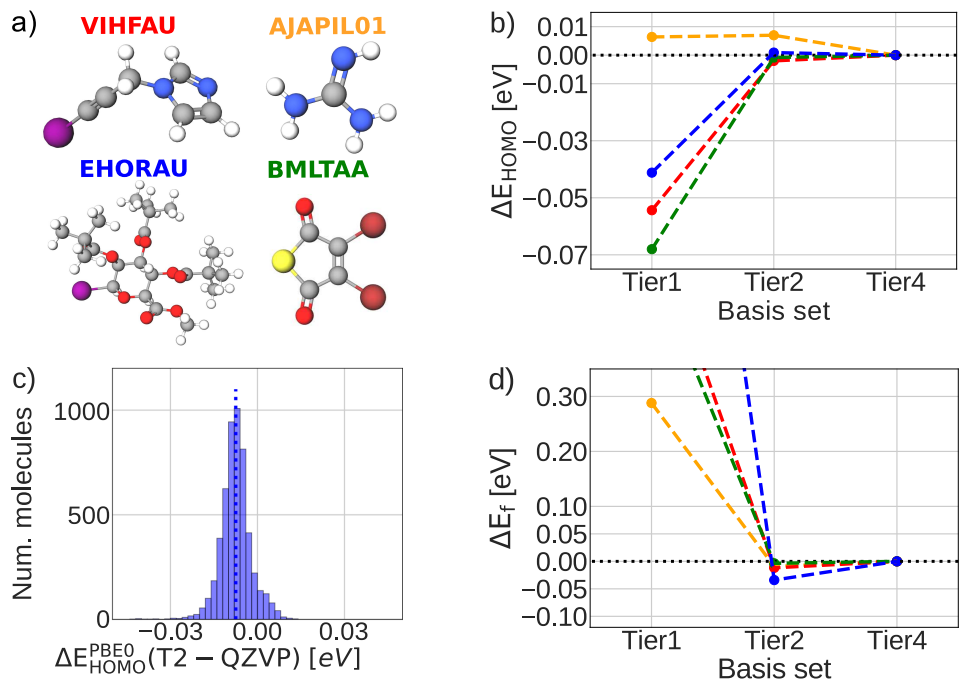}
\end{center}
    \caption{Accuracy assessment of HOMO- and atomization energies computed at the PBE0 (vacuum) DFT level of theory. a) Four example molecules and their \texttt{refcode\_csd} identifiers. b) For the example molecules, the HOMO energy convergence of the Tier1 and Tier2 basis sets is compared against the Tier4 basis set provided with FHI-aims, always employing tight integration settings. c) Difference in HOMO-energy between the Tier2 (T2) and QZVP basis sets for all molecules of the 5k set. The distribution-median is given by a dotted line, located at -0.008 eV. d) Same as b), but for atomization energies $\rm{E_f}$.}

    \label{fig:benchmarks}
\end{figure}
\end{center}

For PBE and PBE0 calculations, the Tier2 basis set of FHI-aims typically provides converged results for both the atomization energy as well as for molecular orbital energies \cite{Jensen2017JpcLett, Lejaeghere2016Science}. The Tier2 basis set has also been used in other large molecular datasets \cite{rupp20012_qm7, Ropo_2016_amino_acids}. We here illustrate the convergence for four selected cases featured in Figure~\ref{fig:benchmarks}a) for PBE0 vacuum calculations at tight settings. As expected, HOMO energies at the Tier2 level are well-converged, here estimated within 0.01 eV around reference values obtained with the largest standard basis set included in FHI-aims (Tier4), see Figure \ref{fig:benchmarks}b). The lower lying orbital energies exhibit a similar convergence behavior (not shown). 

A further quality assessment of predicted HOMO-energies comes from the comparison of Tier2 and QZVP basis set results, as contained in the 5k subset, see Figure \ref{fig:benchmarks} c). We find only a small RMSE of 0.009 eV between the Tier2 and the much larger QZVP basis sets. Figure~\ref{fig:benchmarks} also shows the convergence of the atomization energy of the four molecules in panel d). Again, at the Tier2 level convergence to better than 0.1~eV with respect to Tier4 is observed. This is consistent with results found in a  previous benchmark study \cite{Jensen2017JpcLett}. 

\subsection*{\label{sec:validationGW}Validation of $\boldsymbol{G_0W_0}$ quasiparticle energies}

\begin{figure}[t]      
\centerline{\includegraphics[width=\textwidth]{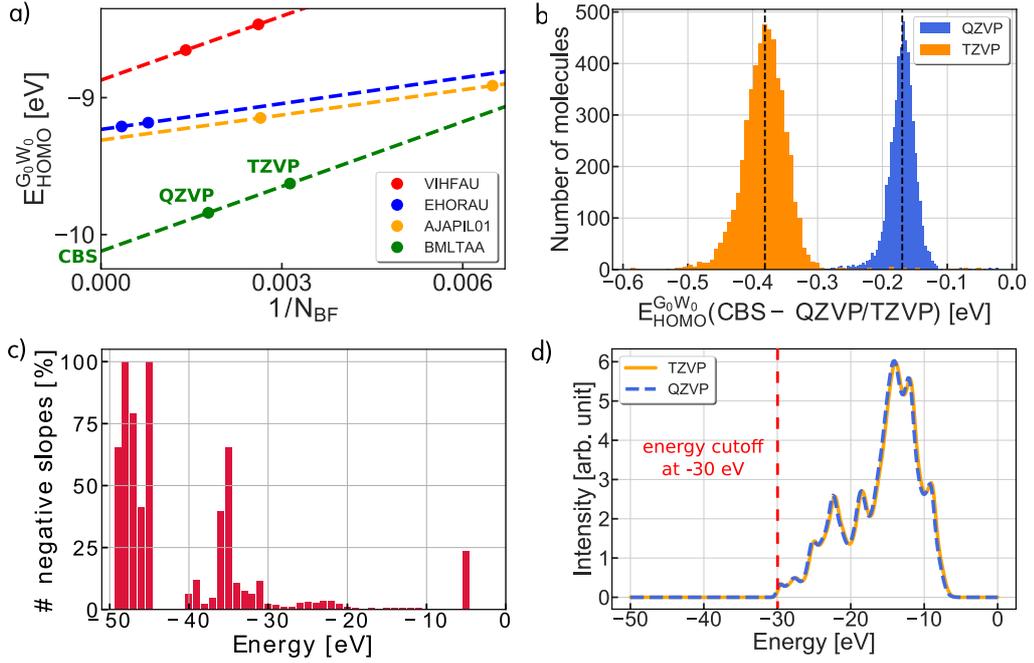}} 
    \caption{Accuracy assessment of $G_0W_0$ quasiparticle energies. a) Convergence of the HOMO $G_0W_0$ energies with respect to the inverse of the number of basis functions $N_\textnormal{BF}$ for the four example molecules shown in Figure \ref{fig:benchmarks}. Dashed lines represent linear straight line fits using the def2-QZVP and def2-TZVP points. The intersection of the straight line with the ordinate gives an estimate for the complete basis set limit (CBS) as indicated for \texttt{BMLTAA}. b) Deviation of the HOMO $G_0W_0$ energies from the CBS limit for the 5k subset. Median values of the distributions are indicated by black dashed lines. c) Percentage of states with negative slope of the CBS fit. d) Average $G_0W_0$@PBE0 quasiparticle spectrum, where each energy state was artificially broadened by a Gaussian distribution.} 
\label{fig:cbs}
\end{figure}

Figure~\ref{fig:cbs}a) shows the convergence of the $G_0W_0$@PBE0 quasi-particle energies with respect to basis set size and their extrapolation to the CBS limit for the four molecules displayed in Figure \ref{fig:benchmarks}a). In all four cases, the $G_0W_0$ energies are not converged even with the largest basis set and CBS extrapolation is required. The slow convergence is typical for the whole 5k set, as demonstrated in Figure~\ref{fig:cbs}b), which reports the deviation of the HOMO $G_0W_0$ energies computed at the TZVP and QZVP level from the CBS limit for all molecules of the 5k subset. The distributions displayed in Figure~\ref{fig:cbs}b) are centered around -0.38 eV (TZVP) and -0.17 eV (QZVP) with a standard deviation of 0.02 eV (TZVP) and 0.01 eV (QZVP) from the median values. Similar results are obtained by including all occupied states above -30~eV in the analysis. In this case, the median value amounts to -0.35 eV for TZVP and -0.15 eV for QZVP. Respective distributions for the deviations of all occupied states from the CBS limit can be found in the Supporting Information. 

The quasiparticle energies at the QZVP level are typically lower in energy than the TZVP values, i.e., the straight line determined from the linear extrapolation to the CBS limit has a positive slope, see Figure~\ref{fig:cbs}a). This empirical observation was already made in the GW100 benchmark study~\cite{Setten2015} for the HOMO level and we also observed it here in our GW5000 study for the valence states. There is no proof that for a given basis set the slope has to be positive. In fact,  for $\sim$ 4\% of the energies level above -30~eV we find negative slopes, as shown in Figure~\ref{fig:cbs}c). This percentage increases considerably in the semi-core energy region between -50 and -30~eV. Such an increase is indicative of either 1) a failure of the analytic continuation used to continue the $G_0W_0$ self-energy from imaginary- to the real-frequency axis  or 2) the insufficiency of the def2-TZVP basis set to converge the deeper occupied states at the DFT level. Based on our analysis in Figure~\ref{fig:cbs}c), we therefore include only states with energies larger than -30~eV in the 5k set. Figure \ref{fig:cbs}d) confirms that the spectral weight averaged over the whole 5k subset is located mostly between -30 to -5~eV and thus, not much spectral information is lost by setting the cutoff threshold to -30~eV.\par

$G_0W_0$ calculations were initially run for 5,500 structures randomly drawn from the 31k set. From these 5,500 molecules, we filtered out molecules for which the analytic continuation of the $G_0W_0$ self-energy is inaccurate or breaks down completely. In FHI-aims the quasiparticle equation is solved iteratively to determine the quasiparticle energies. For some molecules, the pole structure of the self-energy gives rise to multiple solutions and the iterative solution does not converge. We excluded all molecules from the dataset for which at least one TZVP or QZVP level did not converge. Moreover, large differences between the TZVP and QZVP quasiparticle energies are an indication of further problems in the $G_0W_0$ calculation, since the median difference between TZVP and QZVP is only 0.21~eV (see Figure~\ref{fig:cbs}b)). We thus excluded molecules for which at least one level exceeded QZVP/TZVP difference of 0.8~eV. This leaves 5,239 molecules in the 5k set.

\section*{Acknowledgements}

CK, KR and HO acknowledge support from the Solar Technologies Go Hybrid initiative of the State of Bavaria and the Leibniz Supercomputing Centre for high-performance computing time at the SuperMUC facility. CK further acknowledges support by Deutsche Forschungsgemeinschaft (DFG) through the TUM International Graduate School of Science and Engineering (IGSSE), GSC 81. AS, DG, MT and PR gratefully acknowledge computing resources from the Aalto Science-IT project and the CSC Grand Challenge project. DG acknowledges support by the Academy of Finland through grant no. 316168. AS acknowledges support by the Magnus Ehrnrooth Foundation and the Finnish Cultural Foundation. This project has received funding from the European Union's Horizon 2020 research and Innovation Programme under grant agreement No 676580 with The Novel Materials Discovery (NOMAD) Laboratory, a European Center of Excellence. This work was furthermore supported by the Academy of Finland through its Centres of Excellence Programme 2015-2017 under project number 284621 as well as its Key Project Funding scheme under project number 305632. Further support was received by the Artificial Intelligence in Physical Sciences and Engineering scheme (project number 316601).

\section*{Author Contributions}
A.S. and C.K. curated the data, carried out the calculations and postprocessed the results. C.K. performed the calculations at the DFT-levels of theory. A.S. and D. G. conducted the calculations at the $G_0W_0$ level of theory. A.S., C.K., D.G. and J.M. validated the calculations. J.M. analyzed correlations between DFT- and $G_0W_0$-results. M.T., K.R., P.R. and H.O. conceived the original idea and designed the study. All authors cowrote the manuscript.

\section*{Competing Financial Interests}

The authors declare no competing financial interests.

\bibliographystyle{plain}
\bibliography{references}

\end{document}